# Interfacial Properties of Composites Based on h-BN and c-BN in Function of Temperature: a Molecular Dynamics Study

Pedro Altero Parra[1] and Eliezer Fernando Oliveira[1]

[1]São Paulo State University (Unesp), School of Sciences, Department of Physics and Meteorology, Bauru, SP, 17033-360, Brazil.

**Abstract:**

Using molecular dynamics simulations and the ReaxFF force field, we studied a composite based on cubic (c-BN) and hexagonal (h-BN) boron nitride subjected to different temperatures to verify the possibility of a c-BN→h-BN phase transition. Our results demonstrate that the surface termination of c-BN (whether B- or N-terminated) is a crucial factor in the phase transition. The B-terminated c-BN surface presents a lower potential energy than the N-terminated one. However, compared to the potential energy of h-BN, the B-(N-) terminated c-BN surface has a lower (higher) potential energy than h-BN. As the temperature increases, the potential energy of the B-terminated c-BN surface gradually approaches that of h-BN, leading to the beginning of detachment into an h-BN layer around 700 K. With further temperature increase, free h-BN layers can form, which will modify the properties of the composite.

## 1. Introduction

Boron Nitride (BN) is a material with potential for use in several technological applications due to its unprecedented structural, chemical, thermal, mechanical, optical and electrical properties [1]. Among its various polymorphs, those with a two-dimensional hexagonal (h-BN) and three-dimensional cubic (c-BN) crystal structure have been the most studied ones [2-4].

c-BN is a material with several remarkable properties such as hardness, thermal stability, thermal conductivity, electrical insulation, chemical inertness, and optical transparency [4]. These properties make cubic boron nitride an extremely versatile and valuable material in several applications, such as electronic devices, high-performance cutting tools, wear-resistant coatings, and optical components [5]. h-BN has a crystalline structure similar to graphene and presents a laminar structure, high thermal stability, electrical insulation, high thermal conductivity in the in-plane direction, chemical inertness and optical transparency [6]. Such properties make h-BN a versatile material in a variety of applications, including solid lubricants, protective coatings, electrical and thermal insulators, electronic devices, optical components, and others [7].

Recent research has demonstrated that blending c-BN and h-BN could produce new composites with desirable properties for optoelectronics and thermal energy management applications [8,9]. However, there is still a need for a better understanding of the interfacial interaction between the cubic and hexagonal phases, which can influence the mechanical, thermal and electrical properties of the resulting composite [9]. Furthermore, it is known that temperature is a crucial factor for the c-BN→h-BN phase transition to occur [10-11], but it is not known whether the coexistence of both phases can alter this behavior. The abrupt temperature variations that the composite may undergo in certain applications can impact the proportion between the different phases in relation to the initial one, altering the properties of the composite.

In this way, using molecular dynamics simulations, we conducted a theoretical study to further provide information about the behavior of the c-BN/h-BN composite subjected to different temperatures, more specifically at the interface between the different phases. We aim to obtain relevant data on the influence of temperature on the structural conformation of the interface between c-BN and h-BN which can guide experimental development and direct technological applications.

**2. Materials and Methods:**

To conduct this study, we performed Molecular Dynamics (MD) simulations [12]. As we intend to evaluate the physical and chemical interactions at the interface between c-BN and h-BN, as well as the changes in these interactions as a function of temperature, we use the reactive force field ReaxFF [13,14].

The c-BN/h-BN model was constructed using two basic unit cells, one for the c-BN (48 atoms, lattice parameters x = 5.128 Å, y = 4.440 Å, z = 12.560 Å, composed of six bilayers of c-BN) and another for h-BN (48 atoms, lattice parameters x = 5.028 Å, y = 4.340 Å, z = 32.999 Å, six layers). We joined them to generate a unit cell for the c-BN/h-BN composite, leaving a distance of 3.0 Å at the interface between c-BN and h-BN. Then, we replicated it 8x8 to create the supercell of the c-BN/h-BN composite. In our model, h-BN and c-BN are oriented in the [0001] and [111] crystallographic directions, respectively. In Figure 1 it is presented the super cell model for the c-BN/h-BN (6144 atoms, 3072 in h-BN and 3072 in c-BN).

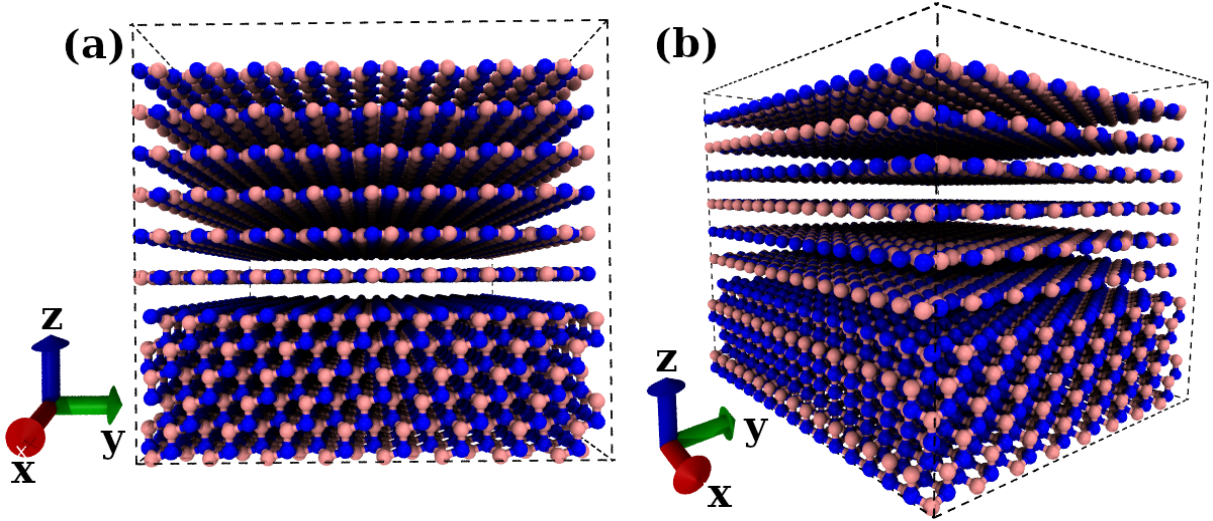

**Fig. 1**. Supercell of the h-BN/c-BN composite in (a) front view and (b) side view.

To study the behavior of the c-BN/h-BN at different temperatures, we subjected our models to four temperatures: 300, 500, 700, and 1000 K. To present our results, we divided the c-BN/h-BN model into chunks along the z direction, each with a thickness of approximately 2.75 Å, in order to obtain the average properties per atom within each chunk. Each chunk contains 512 atoms (256 N and 256 B), resulting in a total of 12 chunks. All measurements were taken after ensuring that the system reached thermal equilibrium at each temperature and the data were averaged over the last 100,000 simulation steps. During the simulations, periodic boundary conditions (PBC) were applied to mimic an infinite composite and to examine interactions between the B-terminated and N-terminated c-BN surfaces in contact with h-BN.

All MD simulations were performed using the Large-scale Atomic/Molecular Massively Parallel Simulator (LAMMPS) software [15] in a NPT ensemble at 0 atm pressure with a timestep of 0.1 fs. The results visualization was conducted with the Visual Molecular Dynamics (VMD) software [16].

**3. Results**

In Figure 2(a), we present the data for the average potential energy per atom of c-BN/h-BN in each chunk at different temperatures. Since the c-BN/h-BN was simulated using periodic boundary conditions (PBC), we obtained information from two interfaces: the B-terminated and N-terminated c-BN surfaces in contact with h-BN. To better visualize the data from both interfaces, we shifted the data in Figure 2(a) so that the chunks numbered 4 to 9 represent data for c-BN, while 1 to 3 and 10 to 12 represent data for h-BN. Thus, the B-terminated c-BN/h-BN interface is located between chunks 3 (h-BN) and 4 (c-BN), and the N-terminated c-BN/h-BN interface is located between chunks 9 (h-BN) and 10 (c-BN). In Figure 2(b), we show the final simulation frame of c-BN/h-BN at different temperatures, colored according to the average potential energy of each atom using an RGB scheme. The color scale ranges from the lowest (red) to the highest (blue) potential energy values.

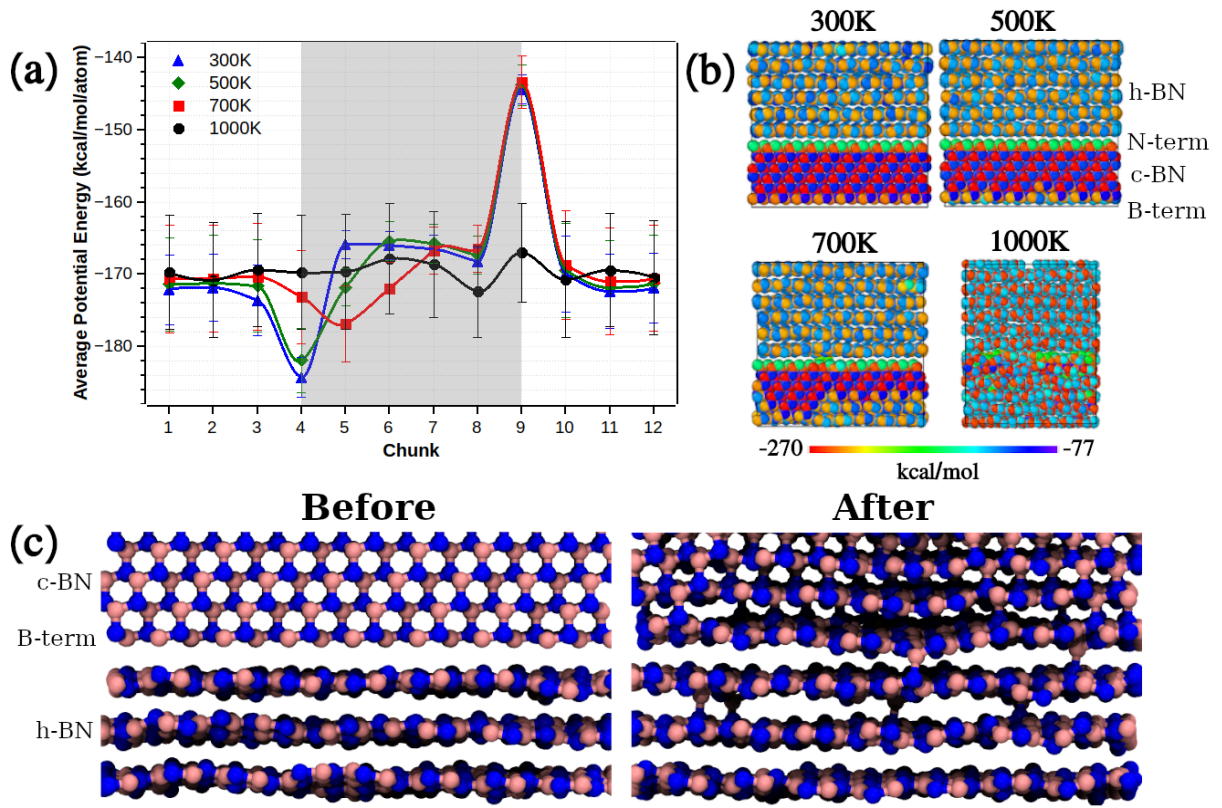

**Fig. 2.** (a) Average potential energy per atom for the c-BN/h-BN as a function of chunk for temperatures of 300, 500, 700, and 1000 K. (b) Final simulation frame of c-BN/h-BN at different temperatures, colored according to the average potential energy of each atom using an RGB scheme. (c) B-terminated c-BN surface before and after it begins to detach from the bulk c-BN at 700 K.

As shown in Figure 2(a), the average potential energy per atom varies with temperature in the interface regions. For temperatures between 300 K and 1000 K, the B-terminated c-BN surface (chunk 4) shows a lower average potential energy per atom compared to the N-terminated c-BN surface (chunk 9). This energy difference becomes more evident in Figure 2(b), where the colors indicate lower levels of potential energy on the B-terminated c-BN surface (bottom surface) compared to the N-terminated c-BN surface (top surface). Compared to the average potential energy of h-BN (chunks 1 to 3 and 10 to 12), the B-terminated c-BN surface (chunk 4) exhibits a lower potential energy level from 300K to 700K; however, as the temperature increases above 700K, its average potential energy tends to approach that of h-BN. In contrast, the average potential energy of the N-terminated c-BN surface (chunk 9) is consistently higher than that of h-BN, regardless of temperature. The average potential energy of h-BN remains approximately constant with temperature. At 1000K, the average potential energies of c-BN and h-BN become indistinguishable, suggesting that c-BN may be fully converted into h-BN layers.

From Figure 2(a), we can infer that the stability of the cubic phase is maintained up to temperatures of approximately 500–700 K, at which point the B-terminated c-BN surface begins to detach from the bulk c-BN, leading to a transition to a layer of h-BN (see Figure 2(c)). Between 500 and 700 K in Figure 2(a), the potential energy of the B-terminated c-BN surface (chunk 4) continuously approaches that of h-BN, suggesting that this surface tends to separate as it aligns with the average potential energy of h-BN. In other words, as the temperature increases, it becomes energetically favorable for the B-terminated c-BN surface to transition to the hexagonal phase, rather than remain in the cubic one. This process suggests a temperature-dependent phase transition, which could significantly alter the properties of the composite in technological applications.

In Figures 3(a) and 3(b), we present the interfacial distance between chunks 3 and 4 and chunks 9 and 10 (representing the interfaces between h-BN and c-BN at both terminations) and the average potential energy difference between them as a function of temperature. At 300 K (see Figure 3(a)), the separation between chunks 9 and 10 (~3.35 Å) is greater than that between chunks 3 and 4 (~2.1 Å). As the temperature increases, the distance between chunks 3 and 4 (B-terminated c-BN surface and h-BN) remains approximately constant up to 700 K, then increases to ~3.5 Å. This increase may be related to the detachment of the B-terminated c-BN surface, transitioning into an h-BN layer, as the experimental interlayer distance between h-BN layers is approximately 3.3 to 3.8 Å [17]. In contrast, the separation between chunks 9 and 10 (N-terminated c-BN surface and h-BN) increases to ~3.5 Å from 500 K to 700 K, suggesting that some structural changes occur at this interface.

In Figure 3(b), the average potential energy difference between chunks 9 and 10 remains constant up to 700 K, then decreases linearly at 1000 K, approaching zero. The average potential energy difference between chunks 3 and 4 begins to decrease at 500 K, reaching approximately -3.0 kcal/mol, and approaches zero at 1000 K. This supports the hypothesis that increasing temperature favors the phase transition of the B-terminated c-BN surface into h-BN layers.

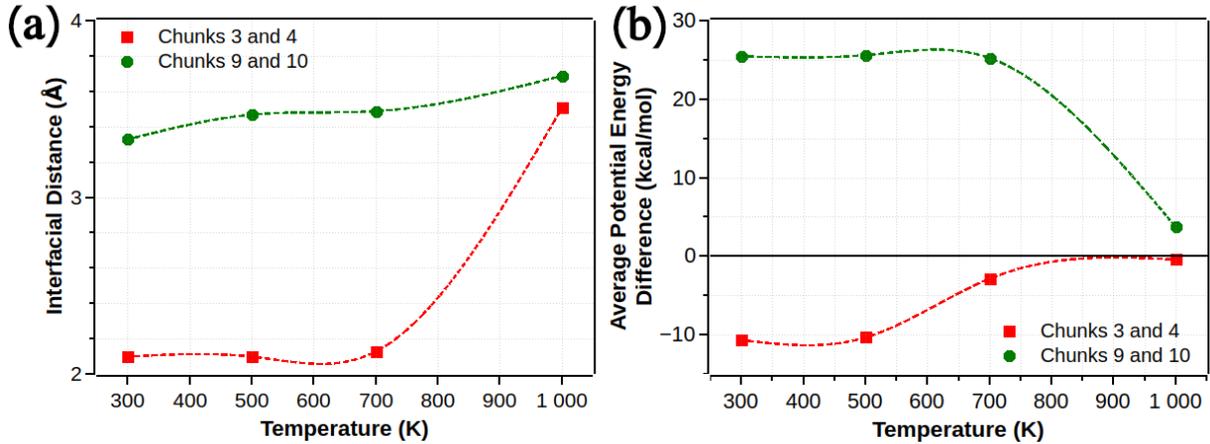

**Fig. 3.** (a) Interfacial distance between chunks 3 and 4 and chunks 9 and 10 (representing the interfaces between h-BN and c-BN at both terminations) as a function of temperature. (b) Average potential energy difference between chunks 3 and 4 and chunks 9 and 10, also as a function of temperature.

## 4. Discussion

Our results indicate that the structural stability of the c-BN/h-BN composite with respect to temperature is strongly influenced by the nature of the c-BN surface termination (B-terminated or N-terminated). The structural stability is maintained up to approximately 700 K, but temperatures above this threshold trigger the transition of the B-terminated c-BN surface to the hexagonal phase (c-BN→h-BN phase transition). This suggests that the proportion of each phase in the c-BN/h-BN composite may change, with an increase in the hexagonal phase. At higher temperatures, it becomes unlikely for the composite to retain its mechanical, electronic, and chemical properties, which will impact the performance of devices based on c-BN/h-BN composites.

## 5. Acknowledgements

Acknowledgement is given to the Center for Scientific Computing (NCC/GridUNESP) of São Paulo State University (UNESP) for its computational support.

## 6. Conflict of Interest

The authors declare no conflict of interest.

# 7. References

[1] S. Roy, X. Zhang, A. B. Puthirath, A. Meiyazhagan, et al. Adv. Mater. (2021) https://doi.org/10.1002/adma.202101589.

[2] K. Zhang, Y Feng, F. Wang, Z. Yang, J. Wang, J. Mater. Chem. C (2017) https://doi.org/10.1039/C7TC04300G.

[3] L. Vel, G. Demazeau, J. Etourneau, J. Mater. Sci. Eng. B (1991) https://doi.org/10.1016/0921-5107(91)90121-B.

[4] S. V. Sominsky, A. M. Zaitsev, J. Mater. Res. Technol. (1999) https://doi.org/10.1016/j.jmrt.2013.03.004

[5] D. Mari, V. Sarin, L. Miguel, C. E. Nebel, Comprehensive Hard Materials, 1st ed. (Elsevier, 2014).

[6] S. Castelleto, F. A. Inam, S. Sato, A. Boretti, Beilstein J. Nanotechnol. (2020) https://doi.org/10.3762/bjnano.11.61

[7] B. Podgornik, T. Kosec, A. Kocijan, C. Donik, Tribol. Int. (2015) https://doi.org/10.1016/j.triboint.2014.09.011

[8] S. Guerini, R. H. Miwa, T. M. Schmidt, P. Piquini, Diam. Relat. Mater. (2008) https://doi.org/10.1016/j.diamond.2008.05.001.

[9] A. Biswas, R. Xu, J. Christiansen-Salameh, E. Jeong, et al. Nano Lett. (2023) https://doi.org/10.1021/acs.nanolett.3c01537.

[10] T. E. Mousang, J. E. Lowther, J. Phys. Chem. Solids (2002) https://doi.org/10.1016/S0022-3697(00)00254-7.

[11] C. Cazorla, T. Gould, Sci. Adv. (2019) https://doi.org/10.1126/sciadv.aau5832.

[12] D. Frenkel, B. Smit, Understanding molecular simulation: from algorithms to applications 1st ed. (Academic Press, San Diego, 2002).

[13] A. C. T. van Duin, S. Dasgupta, F. Lorant, W. A. Goddard, J. Phys. Chem. A. https://doi.org/10.1021/jp004368u

[14] S. J. Pai, B. C. Yeo, S. S. Han, Phys. Chem. Chem. Phys. (2016). https://doi.org/10.1039/C5CP05486A

[15] A. P. Thompson, H. M. Aktulga, R. Berger, D. S. Bolintineanu, et al. Comput. Phys. Commun. (2022). https://doi.org/10.1016/j.cpc.2021.108171.

[16] W. Humphrey, A. Dalke, K. J. Schulten, Mol. Graph. (1996). https://doi.org/10.1016/0263-7855(96)00018-5.

## 8. Authors Contributions

All authors contributed equally to the development of this work and to the preparation of the manuscript.

## 9. Funding

The authors thank the São Paulo Research Foundation (FAPESP) (grant 2023/08122-0), CNPq – National Council for Scientific and Technological Development – Brazil (process 304957/2023-2) and PIBIC-RT - Unesp Institutional Scientific Initiation Scholarship Program (process 9164) for financial support.

## 10. Data availability

Data available upon request.